\documentclass[sigconf]{acmart}

\usepackage{multirow}
\usepackage{threeparttable}
\usepackage{makecell}
\usepackage{soul}

\begin{document}

\title{Toward an Understanding of Developer Behaviour while Using Bug Localization Tools}

\author{Pablo Diaz Pedreira}
\email{pablo.diaz-pedreira@open.ac.uk}
\orcid{0009-0009-0927-9865}
\affiliation{
    \institution{The Open University}
    \streetaddress{Walton Hall}
    \city{Milton Keynes}
    \state{Buckinghamshire}
    \country{UK}
    \postcode{MK7 6AA}
}

\author{Tamara Lopez}
\email{tamara.lopez@open.ac.uk}
\orcid{0000-0001-8766-1896}
\affiliation{
    \institution{The Open University}
    \streetaddress{Walton Hall}
    \city{Milton Keynes}
    \state{Buckinghamshire}
    \country{UK}
    \postcode{MK7 6AA}
}

\author{Michel Wermelinger}
\email{michel.wermelinger@open.ac.uk}
\orcid{0000-0002-6467-3293}
\affiliation{
    \institution{The Open University}
    \streetaddress{Walton Hall}
    \city{Milton Keynes}
    \state{Buckinghamshire}
    \country{UK}
    \postcode{MK7 6AA}
}

\begin{abstract}
    Bug fixing is a complex and time-consuming task in software development.
    Bug localization research tends to focus on the accuracy of automated tools that suggest source code files for developers to look at.
    However, little is known about how developers use these tools in practice.
    This paper reports on an ongoing qualitative user study.
    Eleven participants worked through four realistic bug localization tasks in a controlled environment and were given varying levels of support information offered by a specialized tool.
    Participants were asked to think aloud in a semi-structured interview session.
    The preliminary findings provide insight into three aspects of practice: how developers interact with tools, the role social and contextual information plays, and problem solving.
    The study demonstrates that bug localization is complex and suggests that the adoption of effective tools depends on more than their accuracy.
\end{abstract}


\acmConference[EASE 2026]{The 30th International Conference on Evaluation and Assessment in Software Engineering}{Tue 9 - Fri 12 June 2026}{Glasgow, United Kingdom}


\maketitle

\section{Introduction}\label{sec:introduction}

When an error is found in a software system, it is typically recorded in the form of a bug report (BR), whose main body
contains the description of the problem ~\cite{Lam17, Hindle19, Zimmermann10}.
Later, the BR is triaged and a developer starts to fix the error.
Among the first tasks, the developer must determine where in the source code the changes need to be done.
This is called bug localization (BL) and has been identified as the most time-consuming part of the bug-fixing process~\cite{Hirsch21}.

Automating BL has been extensively studied using information retrieval~\cite{Zhou12, Wang16, Chaparro19} and machine learning \cite{Luo23, Xiao23, Zhang23}.
Both approaches take one BR and the source code as input, and return a ranked list of recommended files to apply the necessary fixes.

Research has focused on improving the accuracy of BL tools, primarily using new computational techniques and additional inputs.
Accuracy is evaluated through closed bug reports from open source projects where the files fixed for the reported bugs are known~\cite{Xiao17}.
By centring the research on accuracy, developers who use the tool's recommendations to support their BL work have been overlooked.

This work aims to address that gap.
The objective is to collect evidence on how a BL tool is used by developers in a scenario similar to a real use case, and to assess how different information given by the tool affect the developer's behaviour, asking:

\begin{itemize}
  \item How do software developers use BL tool information to find the reported error?
\end{itemize}

To answer this question, a study was performed in which participants undertook four BL tasks, and were asked to think aloud.
Additional questions were asked to help participants explain their problem solving process.
A qualitative analysis of the participants' dialogue and actions is underway.
So far, a preliminary analysis of 11 sessions has identified three patterns related to the usage of tools, the social environment, and problem solving.

Behaviours in each pattern reveal an underlying complexity of locating bugs beyond the simple prediction offered by BL tools, regardless of their accuracy.
The examples include iterative refinements in key source code searches, requests of additional information not present in the BR, and reading the BR more than once to change the bug search approach.
Furthermore, we saw that the output of a BL tool may direct, reinforce, or make the developer reconsider key decisions, but it also may mislead and distract.
\section{Background}\label{sec:related-work}

The BL literature focuses on new automated tools and their accuracy, e.g.~\cite{Zhou12, Wang16, Zhang23}, even to predict whether a ranked list is likely to be effective or not~\cite{Le17}.
While these approaches aim to evaluate new BL techniques or new sources of information, how the tools can help in the broader context of bug fixing and how and whether they are actually used by developers is not clear.

Two surveys among software developers have assessed the perceived value of BL tools in real projects.
One is focused on industry~\cite{Li22}, while the other includes open source developers~\cite{Kochhar16}.
These studies found that over 90\% of developers gave “Essential” and “Worthwhile” ratings when asked for the importance of BL tools~\cite{Kochhar16}, and that 68\% are willing to use them~\cite{Li22}.
This positive response contrasts with the expectations of the tool's accuracy.
Only 12\%~\cite{Kochhar16} and 20\%~\cite{Li22} of participants would be satisfied if given a correct Top-1 recommendation only 20\% of the time.
For the use of a tool to be satisfactory, 3 out of 4 participants expect 75\% of Top-1 results to be accurate in one study ~\cite{Kochhar16} while 4 out of 5 participants of the other study expect 80\% accuracy~\cite{Li22}.
This level of accuracy is still out of reach for BL tools~\cite{Xiao23, Zhang23, Luo23}.

The main reason for the lack of adoption of BL tools among professional developers has been identified as the lack of confidence in the tools~\cite{Li22}, with developers preferring to set breakpoints or print statements to understand the program behaviour.
Overall, the results in prior work show a perceived need for help in locating bugs, but that tools have to be reliable.

\section{Method}\label{sec:method}

The aim of this study is to observe how developers use a BL tool in similar conditions to an actual use case.
Following Braz et al.~\cite{Braz22}, the study uses task-directed sessions in which the participant tries to locate the buggy code related with example BRs.
Participants think aloud while performing the tasks~\cite{Whalley14}, with semi-structured questions asked to gain further insight into their thoughts.
After that, the participant's dialogue and actions were analysed.

\subsection{Study design}\label{subsec:study-design}

The study aims to create a setting similar to a real use case of a BL tool, in order to gather truthful reasoning and actions.
All participants were given the same BRs for the same project (Tomcat), to compare their BL approaches.
We selected Tomcat for 4 reasons.
First, it is used extensively in BL research, making it easier to translate the observations to other works.
Second, it is complex enough for the use of a BL tool to be helpful while not being too overwhelming for the participants, given the time constraints.
Third, it is written in Java, a widely used programming language, that made it easier to recruit participants.
Fourth, some participants might already be familiar with this popular project, allowing us to contrast their BL approach with those who don't know the codebase.

For the tasks, we chose 4 closed BRs of varying difficulty for the tool, measured by its accuracy: a single fixed file, ranked first; two fixed files, one ranked first, the other outside the Top-10; a single fixed file, in the Top-10 but not first; no fixed files in the Top-10.

References to the source code are one of the main strategies used by BL tools~\cite{Zhou12, Wang16, Chaparro19} and by developers~\cite{Dit13}.
To minimize the bias of the tool and the risk that participants would use a single strategy, we chose BRs that refer to source code in different ways:
a stack trace, a code example, a file name, and no direct reference.

In Tomcat, the mean number of fixed files per BR is 2.4 and the median is 1~\cite{Ye14, Xiao17}.
We thus selected BRs with 1 to 3 fixed files.
Table~\ref{tab:br-selection-summary} summarizes the BR selection.

\begin{table}[ht]
\caption{Selected BRs}
\begin{tabular}{@{}lll@{}}
BR id & Fixed files (Rank)                          & Reference type \\ \midrule
54087 & HttpServlet (1)                             & stack trace    \\
54095 & DefaultServlet (1), TestDefaultServlet (21) & none           \\
54124 & AsyncContextImpl (4)                        & code example   \\
54144 & Util (31), Out (498), TestOut (1511)        & file name      
\end{tabular}\label{tab:br-selection-summary}
\end{table}

The BL tool selected is DreamLoc~\cite{Qi22}, which uses machine learning.
We modified DreamLoc to take additional information from a project's GitHub repository and trained it on Tomcat for this study.
The added information are seven numbers about files and BRs: 
recent changes, days since the last bug-related change, total changes,  number of authors, 
lines of code, percentage of lines of code changed, and ratio of number of authors to file size.
The model was trained and tested with disjointed sets of closed BRs. 
The 4 BRs used in the study were selected from the test set, according to the criteria given before, and therefore the model had not seen them.

For each BR, we produced three kinds of information: the recommendation list (RL) of the top-10 files; their confidence scores (CS), as computed by DreamLoc; and their summaries (FS), generated by Claude Sonnet 4.0. 
Each was increasingly introduced from task to task.
To reduce confounding factors about how participants locate bugs, we divided participants evenly into 4 groups, each tackling the BRs in a different order and thus with different associated information
(see Table~\ref{tab:br-distribution-per-task-participant}).

\begin{figure}[ht]
    \centering
    \includegraphics[width=\linewidth]{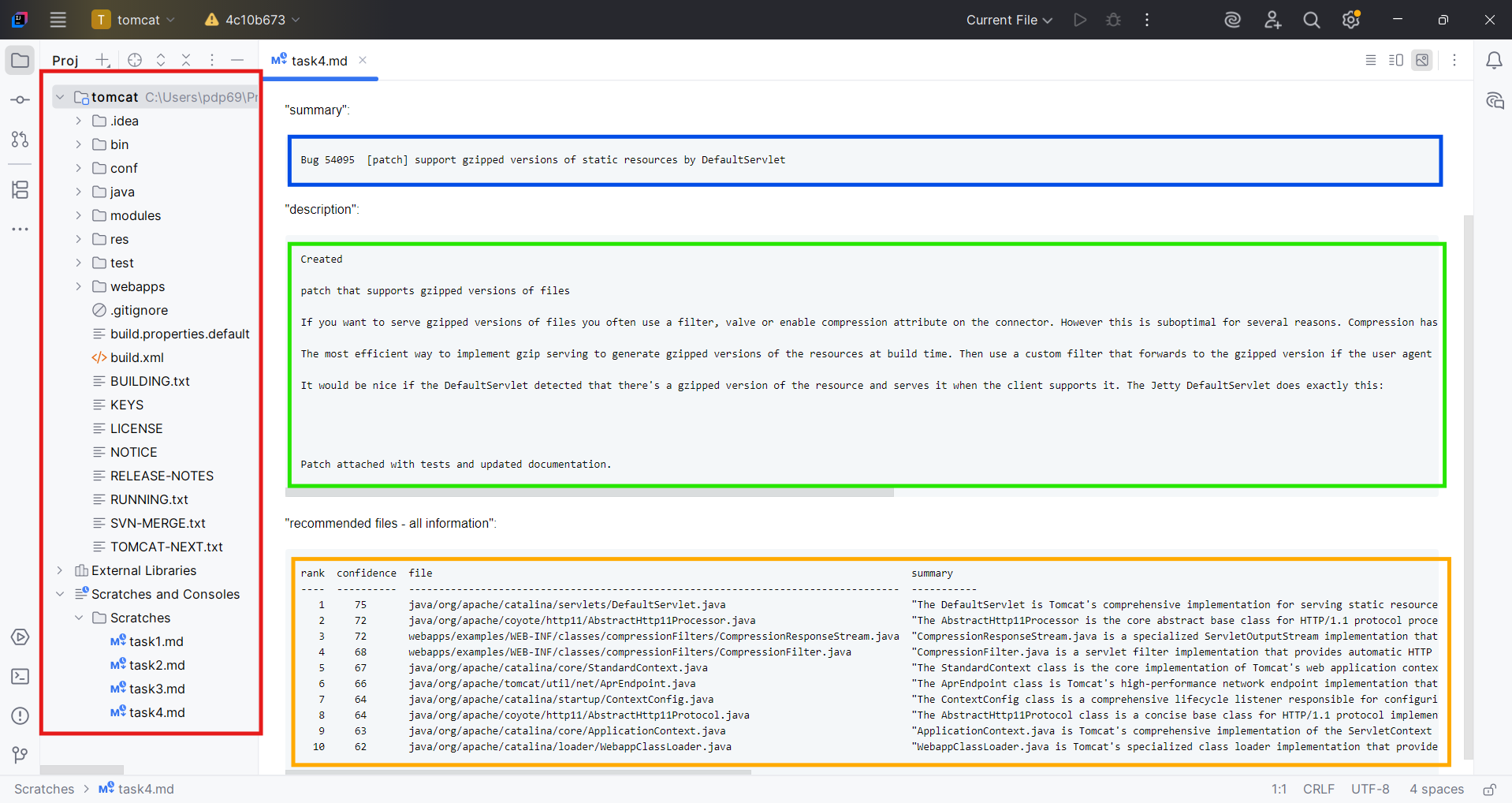}
    \caption{IntelliJ IDE showing BR 54095 for Task 4. Tool information (ranked list, confidence scores, file summaries) in orange. Project explorer in red. BR in blue and green.}
    \label{fig:figure-user_study_taskexample_task4}
\end{figure}

We chose IntelliJ as the IDE for participants to use because it is  commonly used in professional Java projects, and provides two key features for this study: advanced search options and a tree representation of the project structure and files (Figure~\ref{fig:figure-user_study_taskexample_task4}).
To address a potential lack of familiarity with IntelliJ, participants could complete a small introductory practice task.
All but 2 participants skipped the introductory task, marked by an asterisk in Table~\ref{tab:participant-demographic-table}.

\subsection{Participants}\label{subsec:participants}

Sessions were held with 11 participants with different levels of software engineering expertise.
They were not expected to have used BL tools or IntelliJ before.
Contrary to our hopes, no participant was familiar with Tomcat.
Therefore, our setting is more akin to an onboarding scenario where a developer works on a project for the first time.
Table~\ref{tab:participant-demographic-table} shows the demographic information for participants:
the professional software and Java experience are given in years; the bug fixing experience is as participants reported it; the problem-solving confidence level is summarized from their answers; JS stands for Javascript.

\begin{table}[]
\caption{Participants' demographics.}
\label{tab:participant-demographic-table}
\begin{tabular}{@{}llllll@{}}
Part & Prof        & Java        & Bug fixing        & Confident & Other languages \\ \midrule
A1   & 19          & 9           & 7/10              & Very      & C, C++          \\
A2   & 15          & <1          & Daily             & Fairly    & Python, Perl    \\
A3*  & 4           & No          & Only own code     & Very      & Python, C, R    \\
B1*  & 10          & 2--3        & All the time      & Very      & C\#, JS, Python \\
B2   & 2.5         & 3--4        & 1.5 years         & Not       & JS              \\
B3   & 14          & 3           & 7 years           & Very      & C\#             \\
C1   & 10          & No          & 20\%              & Very      & C\#, C++, Python\\
C2   & 4           & <1          & 50\%              & Fairly    & JS, Python      \\
C3   & 28          & No          & 66\%              & Fairly    & JS, SQL         \\
D1   & 2           & No          & Only own code     & Fairly    & Python, Cobol   \\
D2   & 30          & 25          & Daily             & Very      &
\end{tabular}
\end{table}

\subsection{Data collection}\label{subsec:data-gathering}

The sessions were an informal meeting between the first author and the participant.
The tasks were performed exclusively with the researcher's computer, ensuring a common development environment.
Eight sessions were in person; 3 via videoconference, with remote control over the researcher's computer.
The protocol was otherwise identical. 
During the session, the participant's voice and screen were recorded and observation notes were taken.

\begin{table}[ht]
\caption{BR distribution for each task and each participant.}
\begin{tabular}{@{}lllll}
Group (size)& task 1& task 2& task 3& task 4\\
 & RL& RL + CS& RL + FS&RL + CS + FS\\ \midrule
A (3)& 54087  & 54095  & 54124  & 54144    \\
B (3)& 54144  & 54087  & 54095  & 54124    \\
C (3)& 54124  & 54144  & 54087  & 54095    \\
D (2)& 54095  & 54124  & 54144  & 54087    
\end{tabular}\label{tab:br-distribution-per-task-participant}
\end{table}

Each participant was given 4 BL tasks.
For each task, the participant was given a BR, the extra information (Table~\ref{tab:br-distribution-per-task-participant}), and the project's code.
The participant was expected to read the BR and try to locate the buggy part of the code.
Each task was timed for 15 minutes to keep the session's pace and to reduce the risk of not going through the 4 tasks, but participants were free to end the task early or take a bit more time.
Importantly, the study is non-performative: participants were not expected to actually locate nor fix the bugs.

When participants were silent, they were prompted with an open-ended question about their current action, e.g. ``Why did you choose to check this file?''
After finishing each task, the participants were asked to briefly evaluate both the BR and the provided information (recommendation list, confidence score, file summaries).
The demographic questions (Table~\ref{tab:participant-demographic-table}) were only asked after a participant completed all tasks, to reduce researcher bias during the session.
This study was approved by the ethics board of our university.

\subsection{Analysis}\label{subsec:analysis}

Once the data was collected, the audio was transcribed to an anon\-ymised script, and the video to an action log of the participants' actions.
The action log does not reflect the actions' intent, which is covered by the observation notes.

An inductive approach was chosen to provide an expansive view on BL tool usage.
Following ~\cite{Bingham23}, the analysis process began with familiarization: examining the transcript, the action log, and the observation notes.
Here, three patterns were identified, directed by the observation notes.
The patterns correspondingly suggest three different aspects of the problem to be considered: developer interaction with BL tools, social and contextual information, and problem solving.
The three patterns and these three aspects became the focus of the next steps in the analysis presented in this work.

In a second cycle of familiarization, 8 episodes were extracted for detailed analysis.
Each episode represents 3 to 10 minutes of a task.
A third iteration over the episodes, consisted in two levels of coding ~\cite{Bingham23}: attribute codes to organize the data and topic codes to sort the data into searchable categories relevant to the research question.
The following is an example:

\textit{
[Attribute Codes: Transcript (data type); December 2025 (time period); Edinburgh coffee shop (location)]
[Topic Code: BL Tool usage]
45:50 - 46:00
I'm looking for a little one (reading the BL tool file summaries), because it's like a lot of these files size, some of this are huge. Two thousand line files.
}

The codebook, additional details, and examples of the process are included in the supplementary materials.

\section{Findings}\label{sec:findings}

The following subsections describe eight episodes that illustrate how BL tools are used and how they influence developer decision making.
Participant quotes have been normalized when necessary to improve readability, but the full dataset preserved all interjections and repetitions to give indications into reasoning of the participants and to connect thoughts with actions.
The ellipsis (...) is used to indicate a pause and when between square brackets ([...]) indicates parts of the dialogue omitted.

\subsection{Deciding where to start}\label{subsec:deciding-where-to-start}

To locate a bug, developers tend to take a code reference from the BR and search the codebase for it~\cite{Dit13}.
In this study, we observed that searches sometimes are a sequence of iterative refinements, with participants using the tool's results to move through the source code with purpose, rather than in an exploratory manner.

In episodes 1 to 3, participants locate BR 54087, which has a stack trace.
The only fixed file (HttpServlet.java) is the first recommended file and appears in the middle of the stack trace, whereas the first file in the stack trace (Request.java) appears in rank 4.

Episode 1 is taken from task 2 of participant B3: locate BR 54087 given the recommended files and their confidence scores.
B3 comments while reading the BR: ``The stack trace is really useful!''.
B3 follows by checking the BL tool recommendations: ``And if I were to look into the BL tool, we've got the HttpServlet, seems quite relevant. So it reflects what the stack trace requires.''
The participant opens the HttpServlet file, ranked first by the BL tool, and searches within the file first for ``last'', and later for ``getLastModified'', both taken from the BR stack trace.
The BL tool aids the decision to select a file among the stack trace to start the search.
Additionally, it helps reduce the search space, instead of a global search.

Using the BL tool does not always guarantee that suggestions will be followed.
Episode 2 is taken from task 3 of C2: locate BR 54087 using the recommended files and their summaries.
C2 checks the ranked list and the summaries but discards the suggestions.
Their reasoning is:
``Right. Interesting this talking [in the BR] about passing dates and none of the recommended files seem to have anything to do with that. I see an abstract HTTP processor [hovers the mouse over file AbstractHttp11Processor, the 9th recommended file]. But a low level abstract might not be what we necessarily need to be looking at there. So instead I'm going to see if I can open up this section [file Request, the first in the stack trace].''

Next, C2 navigates to file Request and searches for function ``getDateHeader''.
The participant does try to connect the stack trace with the tool recommendations, but discards the suggestions, deeming a low level abstraction file to not be related.
This judgement could be because C2 was not familiar with the project nor Java.

However, the stack trace by itself may have the same directing capability as the ranked list.
Episode 3 is taken from task 4 of participant D2: locate BR 54087 given all information.
D2 jumps to the stack trace, skipping other parts of the BR: ``I got a stack trace, I got an error.''
Next, D2 navigates to the file Request using the project explorer.
Like in episode 2, the participant searches within Request for ``getDateHeader'', the function at the top of the stack trace.
When prompted about why they chose this approach, D2 responded: ``Well, I went straight to the top of this stack. Found this. Found the tests [referring to value checks in the source code].''
The episode shows that the same reduction of the search space takes place (directed local file search instead of global searching).

In these episodes, the tool was used in different ways to decide where to start searching from a stack trace.
First, it supported the correct selection of the file from the stack trace.
Second, the selection made in the stack trace was ultimately discarded, possibly due to a lack of project and Java knowledge.
Third, it was ignored and the person examined the stack trace in order.
In episodes 2 and 3, participants did not have the time to reach the actual buggy file after their initial choice, underscoring the importance of an early search decision.
What separates episode 1 from 2 and 3 is connecting the BR stack trace (where the first file is \emph{not} the buggy file) and the tool suggestions (where the first file \emph{is} the buggy file) before beginning the search.
This allowed B3 to open directly the right file to locate the buggy code.

\subsection{Filling the gaps}\label{subsec:filling-the-gaps}

The previous section showed that the BL tool information has an active role in developer's choices, but the support offered has limitations.
As observed, the BL tool support information sometimes is not enough to complete the localization of the bug, and therefore participants seek additional sources, either formal (like design documentation) or social (like turning for help to the reporter).
The following episodes illustrate how the information from the BL tool needs to be supplemented by other kinds of information.

Episode 4 is taken from task 4 of participant C1: locate BR 54095 given all the information (ranked files, confidence scores, file summaries).
C1 starts out confident about the tool's suggestions and deems all types of support information useful.
However, for the last steps of the localization, they found it necessary to identify more details to continue.
So while the support information was welcome, it fell short and additional documentation was requested.

C1 starts to relate the changes mentioned in the BR with the file recommendations and summaries, and adds: ``That DefaultServlet [rank 1 file], I can read that and learn what that is [reads the file summary].''
The confidence score is then evaluated: ``This confidence [score] seems a bit low [75/100] because it seems 100\% obvious that that's the code that has been changed [refers to the patch mentioned in the BR].''
C1 follows by searching the terms ``gzip'' and ``compression'' in DefaultServlet.
At the end of the task, rather than being confident on where the bug is, C1 explained: ``In real life, I probably spent a lot of time going through the changes...
I'd just roll the git diff to figure out what he's actually done to get an overview of the approach. But it's also to me like an architectural decision, so it might be a case where I was looking at a software engineering design discussion.''
Similar observations were mentioned throughout the task: ``But I also want the bigger picture: Where this fits in, in the architecture of the tool.'', ``You would need the documentation... The architectural design of the system.''

In episodes 5 and 6, the participants only have access to the file recommendation list, and the BL tool is perceived differently.

Episode 5 shows that BL tools are not always enough to guide and support developers on their own, even when accurate. Episode 5 is taken from task 1 of participant D1: locate BR 54095 (the same as in episode 4) given the ranked files.

D1 starts being confused about the BR: I'm saying this description doesn't tell me what I'm meant to do. [...] If I were actually a member of this project, I might know what happens [in the BR].''
The confusion motivates D1 to look at the BL tool information, but seems to not help: ``So that also means that I don't feel able to interpret these file recommendations. I mean, these are a bunch of files, they could be relevant in some way. [...] I could open up one of these files and look at it, but I wouldn't know what I would be looking for, how it related to the task that I meant to perform.''
At the end of the task, the participant was still unclear about the bug and was not able to find any potentially buggy code.
The file recommendation list alone does not seem to help, despite D1 relying on it and the BL tool ranking first the fixed file.

Episode 6 is taken from task 1 of participant B2: locate BR 54144 given the ranked files.
In this episode, the BR lacked context and the tool was not accurate, so B2 would resort to asking the reporter.

B2 has a number of questions about the BR: ``So it's sort of clear, but I'm missing context. [...] Why are they erased now? What problems is this creating along the line? I try to visualise what else it [the bug] is impacting.''
The evaluation of the BL tool by B2 was negative: ``It was not very useful. Having a list of files, while I was trying to locate if there's any of these calls happening in those files...
I could have saved time if I started from the place that it is pointing to [Out file mentioned in the BR].''
The BL tool file recommendations for this BR do not contain any buggy file.
In turn, B2 perceives that the tool output wasted their time.
Nonetheless, B2 managed to make progress in the task, commenting on how they would continue: ``If I read this [the BR], I will have an idea where to start looking like I've done at the end now, but I would have to feed the person that wrote this file with a couple of questions.''

These episodes show that developers rely on contextual information from other sources, both technical and social when finding bugs.
They demonstrate that the information from the tool cannot always fill that gap.
The lack of additional documentation, reading support, and contextual information from the BR is a limitation of current tool approaches.
Confidence scores and file summaries may compensate somewhat for this limitation in the early stages of the localization, as suggested by the contrast in tool perception in episode 4 (full support information) compared to the other two.

\subsection{Reinterpreting available information}\label{subsec:reinterpreting-available-information}

So far, all participants have started the tasks by reading the BR in varying levels of detail.
Occasionally during the task, the participant gets stuck, resorts to reading the BR\@ again, and ends up changing their approach to the task.
In the process, the way participants engage with the BL tool is also affected.
Revisiting the BR clarifies the situation for the participant.
Sometimes the BL tool has led the participant astray, and sometimes the BL tool is the mechanism that gets the participant moving forward again.

Episode 7, like episode 6, is taken from task 1 of participant B2: locate BR 54144 given the ranked list.
In episode 7, after following the tool recommendations, B2 decides to stop checking files and reads the BR again: ``I don't think this has anything to do with that [the BR]. [...] Let me read the task again.''
During the BR revisit, B2 finds a file reference missed in the first read: ``OK, maybe I should start from this one [points to file Out in the BR]. If I can find it [the bug], and it's not too late. [...] The code that the plugin generates calls value.toString(). Let me see if that's true. [opens Out.java] There are any call here? OK, it's there.''
B2 then is able to continue the task reading file Out and searching for ``toString''.
B2 comments at the end of the task: ``I think I've managed to locate now, with the description, where [the BR] was pointing me to and that I've neglected [...] until the end.''

Episode 8 is taken from task 3 of participant B3: locate BR 54095 given the recommended files and their summaries.
B3 seems stuck while reading DefaultServlet, the first ranked file: ``Not entirely sure where [in the source code] it [the BR] is talking about at the moment''.
Then B3 spends 3 minutes looking at the BL tool before deciding to go back to the BR: ``Making sure that I'm... Make sure I understand the problem sufficiently.''
While reading the BR, B3 starts filtering the tool recommendations with the aid of the file summaries: ``Unless the bug is about the samples [files ranked 3 and 4], but it doesn't, it doesn't look that way. I do feel like a DefaultServlet behaviour. Somewhere about the DefaultServer.''
After reading the BR, B3 checks file DefaultServlet, and discards it.
B3 goes back to the tool suggestions and moves to the rank 2 file, AbstractHttp11Processor.java.
There, B3 starts to find parts of the code that match the expectations of where the bug should be.
In reflecting on the BL tool for this task, B3 states: ``First, I didn't find it useful [the BL tool], but in the end helped me find the problem because it led me to the file [AbstractHttp11Processor]. There is no way I would have spotted it otherwise.''
The second reading of the BR seemed important to reframe the tool's suggestions.
File AbstractHttp11Processor is not a buggy file for the BR\@.
Nonetheless, the participant tried searching within DefaultServlet during the task, even before episode 8, and after arriving at AbstractHttp11Processor B3 made progress in their understanding of the bug.
The tool was perceived as beneficial and ultimately allowed B3 to move forward.

BL tool information is a part of the iterative comprehension process that characterizes bug localization.
The additional thoughts, effort, and increase in source code familiarity, after trying to locate the bug for a while, may allow the participants to elicit new information from both the BR and the tool.
This in turn results in following a new direction (episode 7) or reinterpreting parts of the BR and tool information (episode 8).
This means that tool suggestions are not necessarily useful only at one time, presumably at the beginning of the task, but rather they evolve with the information developers gather and understanding they gain about the problem.

\section{Limitations}\label{subsec:limitations}

In bringing together responses from different participants, the findings in this report provide a constrained view on problem solving during bug localization.
Some aspects of the study design may have influenced the interpretation.
For example, though the study was designed to reflect tool use in natural settings, time restrictions may have pushed participants to reject tool output or to change objectives.
In addition, the lack of familiarity with Tomcat, while planned for in the task design, may have affected behaviour.
Finally, the analysis presented here is preliminary, and was primarily performed on a subset of the sessions.

As a task-directed study, time and difficulty constraints cannot be removed.
These limitations were mitigated by making the tasks non-performative and being flexible about time.
The interpretation made here will be checked through deeper engagement with the full dataset.
The study does not cover all possible scenarios, like multiple fixed files in the 10 recommendations shown.
Extra scenarios were discarded after the pilot phase, to avoid participant fatigue~\cite{Rodeghero15}.

The study uses one BL tool and one LLM because we are interested in how users engage with the information such tools produce, rather than evaluating the tools' accuracy.
Using any other BL tool and LLM would similarly support the research aim.

\section{Discussion}\label{sec:discussion}

The preliminary analysis presented here is based on early patterns identified in the data.
The patterns themselves were chosen as a point of departure to provide context for the data analysis and interpretation of the tool interaction.
It is apparent that the developers engage with the tool, and use it in various ways as described in Section~\ref{sec:findings}.

By inspecting a series of episodes, we noted instances of how software developers use BL tool information to find the reported error.
First, by guiding the decision of where to start searching following a stack trace (episodes 1, 2, 4).
Second, by reinforcing the information available in the BR (episodes 1, 4).
Third, by offering an alternative option when the developer gets stuck (episode 8).

We also noted some of the limitations BL tools face in practical scenarios.
They are ignored (episode 3), they are not able to support an already confused developer (episode 5)~\cite{Sillito08}, and are occasionally misleading (episodes 6, 7).

One known problem in BL is the low quality of BRs~\cite{Li22}, partly because not all the information needed is present in them~\cite{Aranda09}.
Episodes 4, 5, and 6 confirm that developers still require additional information even when they use tools.
However, comparing episode 4 with 5 and 6 suggest that additional support information could alleviate this need~\cite{Sillito08}, at least in early stages of bug localization.
Future analysis work may provide additional evidence to support this possibility.

Previous work found that developers require a high level of accuracy for the tools to be satisfactory and therefore adopted~\cite{Kochhar16, Li22}.
However, during the study, no participant stated any problem with the tool's accuracy.
To further support this, in the data analysed so far, there was no engagement with the confidence score (used here as a proxy for accuracy) other than the one in episode 4.
One explanation could be that participants may form an idea of the tool's accuracy during task 1, when they only have access to file paths.
Further data needs to be analysed to support this premise.
In contrast, file summaries were used in episodes 2, 4, 8, and were considered for the participant's reasoning as demonstrated with quotations.
This may suggest that, in contrast to previous evidence based on survey data~\cite{Li22}, the barrier to using BL tools is not strictly low accuracy.

\section{Conclusion}\label{sec:conclusion}

To our knowledge, this research is the first qualitative empirical study to examine how developers engage with bug localization tools.
By observing developers on realistic tasks supported by varying amounts of information (ranked file list, confidence scores, file summaries), we identify how BL tools shape behaviour, and how developers integrate recommendations into their reasoning.
The main contribution gives insight beyond tool accuracy: tool usage itself, social context information, and problem solving challenges.
We document the rational and informational strategies developers employ when interacting with BL tools, highlighting developers’ need for additional sources of information, and showing that BR re-reads help overcome problem solving challenges.

Our findings show that developers do not rely on recommendations alone; instead, they combine BL outputs with purposeful searches and contextualize results through the bug report.
Even features intended to help, such as confidence scores or LLM‑generated summaries, influence behaviour in nuanced ways, sometimes improving navigation but also occasionally misleading.
These findings underscore the importance of BL tools to communicate uncertainty, rationale, and context more effectively.

Future work will expand the analysis, documenting additional strategies developers use with BL tools, how developers relate the BR with the BL tool information and the source code, and refine how multiple sources of information influence developer behaviour.
Ultimately, this research contributes to understanding of tool assisted bug localization practices, identifies aspects of this complex problem space that developers need more help with, and shows the limitations of BL tools in real-world scenarios.

\section*{Supplementary Files}\label{sec:supp_files}
https://doi.org/10.6084/m9.figshare.31449289

\section*{Acknowledgements}
We thank the developers who participated in our study.
We thank the reviewers for their helpful comments.
The research protocol was approved by The Open University Human Research Ethics Committee, ref. 2025-1006-1.
This work was partially supported by project Empirical Data-Driven Bug Localization,
funded by Huawei Technologies (Ireland).

\bibliographystyle{ACM-Reference-Format}
\bibliography{thesis}

\end{document}